\documentclass[floatfix,nofootinbib,superscriptaddress,twocolumn,prl,showpacs,showkeys,preprintnumbers]{revtex4-2}

\usepackage{latexsym,amsmath,amsfonts,amsthm,amssymb,bbm}
\usepackage{epsfig,graphics,color,calc,graphicx,pict2e}
\usepackage{comment} 

\usepackage{mathrsfs}
\usepackage{tikz}
\usepackage{tikz}
\usetikzlibrary{shapes,arrows}
\usetikzlibrary{intersections,matrix, positioning}

\usepackage{todonotes}
\usepackage[printonlyused,withpage]{acronym}
\usepackage[acronym]{glossaries}

\newlength{\pecettawidth}
\begin{document}
\title{\normalsize\Large\bfseries 
Boundary-Driven Anisotropic Coarsening in Conserved
Phase Separation}


\author{Emilio N.M. Cirillo\thanks{emilio.cirillo@uniroma1.it}}
\affiliation{Dipartimento di Scienze di Base e Applicate per l'Ingegneria, 
             Sapienza Universit\`a di Roma, 
             via A.\ Scarpa 16, I--00161, Roma, Italy.}

\author{Nicklas J\"averg\r{a}rd\thanks{nicklas.javergard@kau.se}}
\affiliation{Department of Mathematics and Computer Science, Karlstad University, Karlstad, Sweden,}


\author{Adrian Muntean\thanks{adrian.muntean@kau.se}}
\affiliation{Department of Mathematics and Computer Science, Karlstad University, Karlstad, Sweden,}

\author{Stela Andrea  Muntean\thanks{andrea.muntean@kau.se}}
\affiliation{Department of Engineering and Physics, Karlstad University, Karlstad, Sweden}

\date{\empty} 

\begin{abstract}
Universal scaling in phase separation is typically
assumed to be isotropic in systems with conserved
dynamics. Here we show that boundary forcing alone can
break this dynamical scaling symmetry, 
leading to anisotropic coarsening, with different
effective global growth laws observed parallel and
perpendicular to the boundary.
We consider a
ternary mixture with two conserved components and a
passive species undergoing surface evaporation, which
provides a simple setting to investigate this effect. In
this case, evaporation leads to a progressive mass loss
and to the formation of macroscopic concentration
gradients, which, in turn, drive anisotropic coarsening,
with different effective growth laws observed parallel and
perpendicular to the boundary. At the same time, bulk
regions appear to retain the standard Model B scaling,
suggesting that the observed anisotropy is mainly
induced by boundary fluxes rather than by changes in the
intrinsic dynamics.
Our results indicate that boundary conditions
can play an important role in breaking scaling symmetry
and may offer a way to influence coarsening behavior in
nonequilibrium phase separation.
\end{abstract}


\keywords{Phase separation, ternary mixture, evaporation, morphology formation in 3D, Blume--Capel model,  coupled non-local parabolic system,  Monte Carlo method, finite volume approximations.}

\maketitle

Morphology formation in multicomponent mixtures
\cite{Laradji1996} represents a central problem in
nonequilibrium statistical physics, with relevant
implications for functional materials such as polymer
blends, organic coatings, and thin-film devices
\cite{hoppe2004organic, Moons2002, Moons2007}.
In systems with conserved order parameters, phase
separation typically proceeds via spinodal decomposition
followed by coarsening \cite{CahnHilliard1958}, and is
generally described in terms of universal scaling laws
\cite{Bray1994}. In particular, for isotropic systems
with diffusive dynamics (Model B), the characteristic
domain size is expected to grow as $t^{1/3}$, largely
independently of microscopic details
\cite{HohenbergHalperin1977,bray1990universal,CGRS2023}.
In this framework, morphologies are mainly selected by
composition and interfacial energies, while the scaling
properties remain essentially isotropic.

A fundamental open question concerns how external driving
and boundary conditions may affect these universal
coarsening laws.
While anisotropic effective growth has been
observed in systems with intrinsic directional
interactions, such as competing couplings or applied
shear \cite{Cirillo1997, Corberi1999,Davis2025,Onuki2002},
in those cases the anisotropy is already built into the
microscopic dynamics.
By contrast, much less is known about whether, and to
what extent, purely external forcing acting only at the
boundaries can induce anisotropic scaling in otherwise
isotropic conserved systems.

Here we show that boundary forcing alone can provide a
simple mechanism to induce anisotropic coarsening. We
consider a ternary mixture with two conserved components
and a passive species undergoing surface evaporation
\cite{Deegan1997,Larson2014}.
In these systems, phase separation arises from the
interplay between composition and the nonequilibrium
dynamics associated with solvent evaporation; see e.g.
\cite{Myatt, Morpho_ML,Thiele2010}. 
The resulting
boundary-driven mass loss leads to the formation of
macroscopic concentration gradients, which break the
symmetry at the system level and give rise to
direction-dependent growth laws.
In particular, we observe different scaling exponents
parallel and perpendicular to the evaporation direction,
while bulk regions retain a behavior consistent with
standard Model B. This suggests that the observed
anisotropy is mainly due to boundary-induced fluxes,
rather than to modifications of the intrinsic dynamics.

Boundary effects on spinodal decomposition have also been
studied in the framework of surface-directed spinodal
decomposition, where a surface field favoring one component of a
binary mixture generates surface-directed concentration waves
and time-dependent enrichment and depletion layers normal to the
wall \cite{Puri1992,Bhattacharyya2026}. These effects produce anisotropic
structures close to the surface. By contrast, the anisotropy
reported here is not induced by preferential wetting but by
boundary-driven evaporation and the associated macroscopic
concentration gradients.

We adopt 
a three-state
lattice model 
\cite{cjlmm3D01,Andrea_PhysRevE}
in which the spin variable
$\sigma(i)\in\{-1,0,+1\}$ represents two mutually repelling
species ($\pm1$) and a passive component ($0$) on a cubic
lattice with periodic boundary conditions.
The Blume-Capel Hamiltonian \cite{blume1966} reads
\begin{equation}
H(\sigma)=
J\sum_{\langle i,j\rangle}[\sigma(i)-\sigma(j)]^2 ,
\end{equation}
with $J>0$. Interfaces between unlike active species cost
$4J$, while those involving the passive component cost $J$,
effectively screening repulsion between the active species.

Bulk dynamics is implemented via conserved Kawasaki exchanges
\cite{cjs2024,MBR2024},
ensuring local conservation of the $\pm1$ densities. Evaporation
is introduced at the top boundary by allowing passive sites
to be stochastically converted into active spins with 
probability $\alpha$, thus breaking
total mass conservation while preserving the relative fraction
of the two active components.
In all simulations we will use  $J=1$, 
$L=256$, and inverse temperature $0.80$.

We stress that periodic boundary
conditions are imposed on all three spin states. Evaporation
acts only through the stochastic conversion of passive sites at
the top boundary, while the active species are never directly
created, removed, or exchanged through the boundary. Their
dynamics is entirely governed by Kawasaki exchanges
and is therefore affected by evaporation only indirectly
through the depletion of the passive component.

To complement the lattice description and address
possible discretization artifacts, we also consider
a continuum counterpart obtained from a Kac-type
long-range version of the Hamiltonian
\cite{Marra,lebowitzpenrose1966,GiacominLebowitz1997}.
In the hydrodynamic limit, the system evolves
according to coupled nonlocal drift–diffusion
equations for the magnetization $m(t,x)$ and
the squared spin density $\phi(t,x)$,
\begin{equation}
\left\{
\begin{array}{l}
\partial_t m
= \nabla\!\cdot\!\left[
\nabla m
-2\beta(\phi-m^2)(\nabla J * m)
\right] 
\\
\partial_t \phi = \nabla\!\cdot\!
\left[ \nabla \phi -2\beta m(1-\phi)(\nabla J * m) \right]. 
\end{array} \right. 
\end{equation} 
Here $1-\phi$ represents the passive component concentration. 
Evaporation is incorporated 
through boundary conditions allowing passive component 
loss at the free surface. 
The inequality $|m|\le\phi\le1$ is preserved during the 
evolution \cite{lyons2024phase}. 
Unless otherwise specified, simulations start from random initial conditions 
with prescribed initial fractions of the three components.
In the continuum model evaporation is introduced by means of 
a suitable boundary condition at the top and bottom face of 
the lattice
\begin{equation}
[\nabla \phi - 2\beta m (1-\phi) (\nabla J \star m ) ] \cdot n = \alpha' (1-\phi)
\end{equation}

In the absence of evaporation, we previously reported
morphology formation in both two- and three-dimensional
systems \cite{Andrea_PhysRevE,cjlmm3D01}.
In three dimensions, the characteristic domain size
follows a $t^{1/3}$ growth law. 
Coarsening remains
isotropic and is consistent with conserved Model B
dynamics. Depending on the passive
fraction, the resulting 
morphologies consist either of
interpenetrating active domains separated by thin
passive layers or of isolated active clusters
embedded in a passive background.

The domain length was extracted
from the normalized two-point correlation function
$G(k,t)=\langle\sigma(i)\sigma(i+k)\rangle/
\langle\sigma(i)^2\rangle$.
Correlation functions measured along the three spatial
directions displayed identical scaling behavior,
consistent with isotropic conserved (Model B)
spinodal decomposition.

When evaporation is enabled, global isotropy is
lost. Starting from a configuration rich in the
passive component, active species initially form
isolated clusters. As evaporation proceeds and
the passive fraction decreases, these clusters
merge into intertwined structures.

\begin{figure}
\centering
\includegraphics[width=0.31\columnwidth,height=0.31\columnwidth]{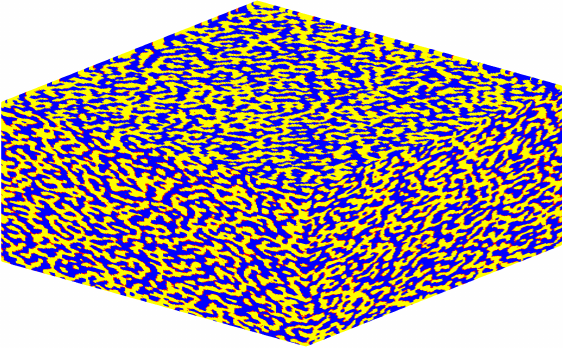}%
\hspace{0.01\columnwidth}%
\includegraphics[width=0.31\columnwidth,height=0.31\columnwidth]{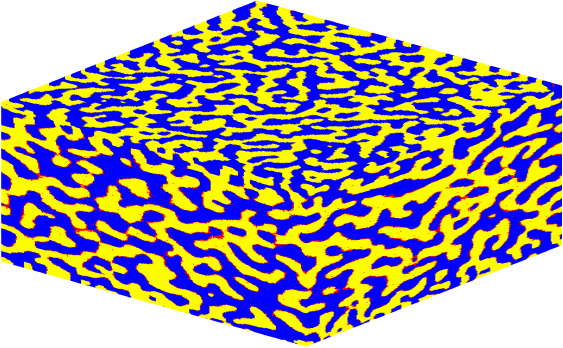}%
\hspace{0.01\columnwidth}%
\includegraphics[width=0.31\columnwidth,height=0.31\columnwidth]{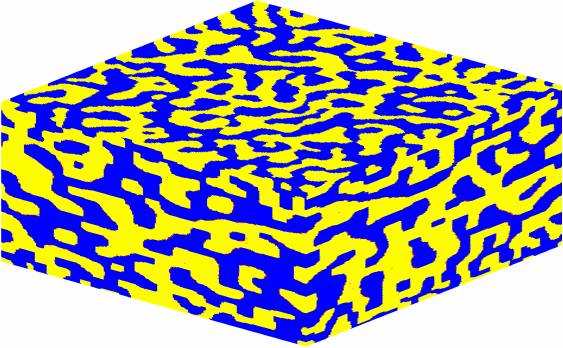}
\\[0.05cm]
\includegraphics[width=0.31\columnwidth,height=0.31\columnwidth]{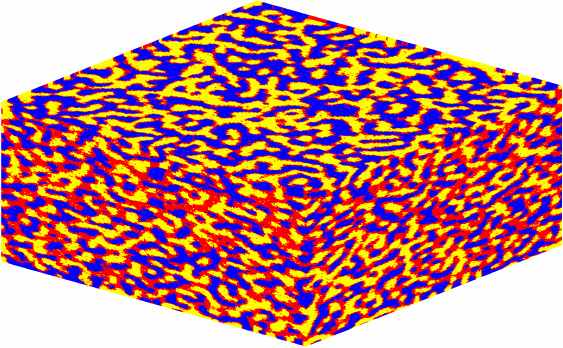}%
\hspace{0.01\columnwidth}%
\includegraphics[width=0.31\columnwidth,height=0.31\columnwidth]{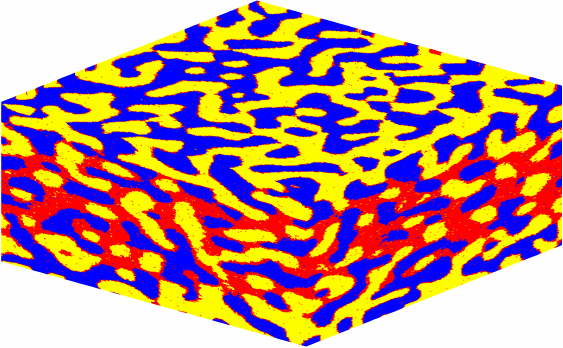}%
\hspace{0.01\columnwidth}%
\includegraphics[width=0.31\columnwidth,height=0.31\columnwidth]{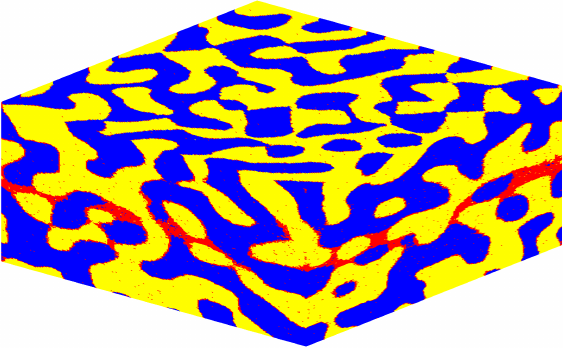}
\\[0.05cm]
\includegraphics[width=0.31\columnwidth,height=0.31\columnwidth]{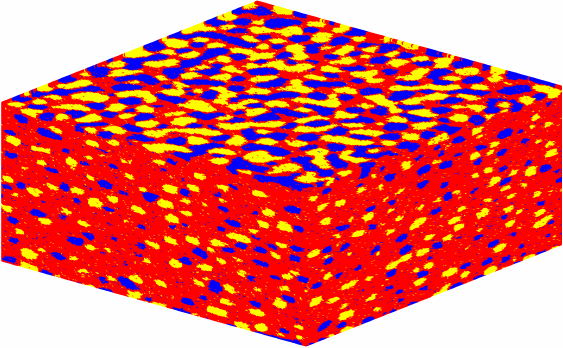}%
\hspace{0.01\columnwidth}%
\includegraphics[width=0.31\columnwidth,height=0.31\columnwidth]{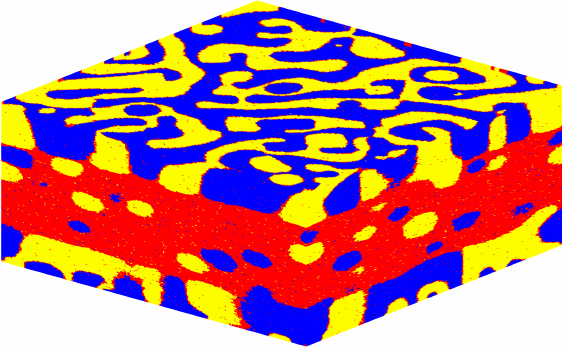}%
\hspace{0.01\columnwidth}%
\includegraphics[width=0.31\columnwidth,height=0.31\columnwidth]{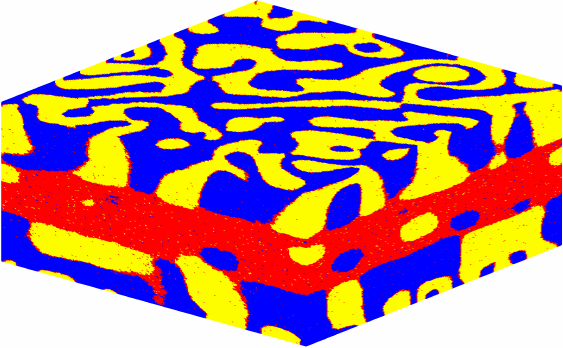}
\\[0.05cm]
\caption{Typical configurations of the lattice model in the presence
of evaporation for $\alpha=10^{-3}$. Top row:
$c_0=0.1$ at $t=10^3$, $10^4$, and $4\times10^4$ (left to right).
Center row: $c_0=0.4$ at $t=10^3$, $10^4$, and $3\times10^4$.
Bottom row: $c_0=0.8$ at $t=10^3$, $2\times10^4$, and $3\times10^4$.
The planes shown correspond to $(x,0,z)$ (left), $(L-1,y,z)$
(right), and $(x,y,L-1)$ (top), with $x,y,z=0,\dots,L-1$ of the
cubic lattice $\Lambda$.}
\label{fig:e256conf_evap0001}
\end{figure}

Figure~\ref{fig:e256conf_evap0001} shows typical lattice configurations
for evaporation probability $\alpha=10^{-3}$. For initial passive
fraction $c_0=0.1$, domain evolution is nearly isotropic, resembling
the case without evaporation, with dynamics slowing as the passive
component is depleted. In all three cases, the ratio of longitudinal
to transverse domain size is approximately $1$.

For higher initial passive fractions ($c_0=0.4$ and $0.8$, center and
bottom panels), domains grow faster along the evaporation direction
(the third spatial direction), while effective 
growth perpendicular to it remains
slower. This anisotropic coarsening is particularly pronounced for
$c_0=0.8$ and moderately visible for $c_0=0.4$. Visual inspection suggests
that the residual passive component remains near the center of the
lattice, and that domains in this region are markedly elongated along
the evaporation direction, revealing strong anisotropic growth driven
by surface mass loss. Quantitatively, for the three times reported in
the figure, the ratio of longitudinal to transverse domain size is
$1$, $1.01$, and $1.34$ for $c_0=0.4$, and $1.02$, $1.59$, and
$1.73$ for $c_0=0.8$.

\begin{figure}
\centering
\includegraphics[width=0.31\columnwidth,height=0.31\columnwidth]{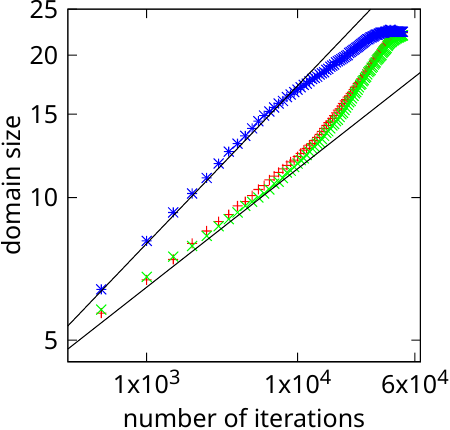}%
\hspace{0.01\columnwidth}%
\includegraphics[width=0.31\columnwidth,height=0.31\columnwidth]{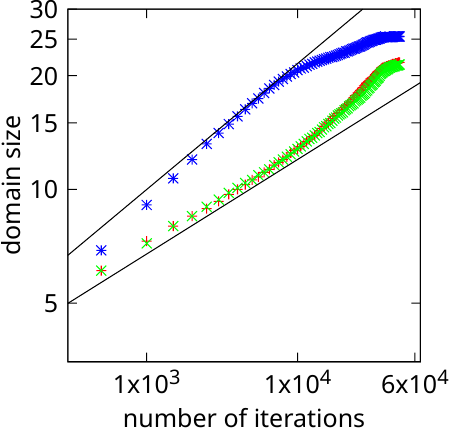}%
\hspace{0.01\columnwidth}%
\includegraphics[width=0.31\columnwidth,height=0.31\columnwidth]{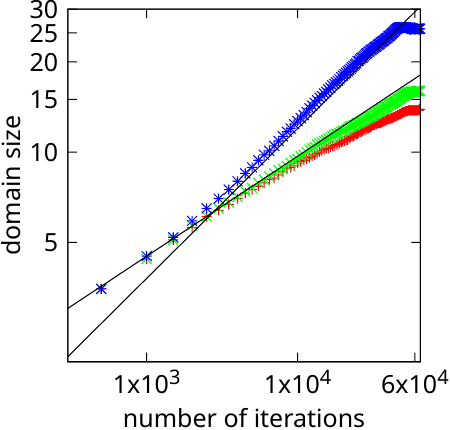}
\\[0.05cm]
\includegraphics[width=0.31\columnwidth,height=0.31\columnwidth]{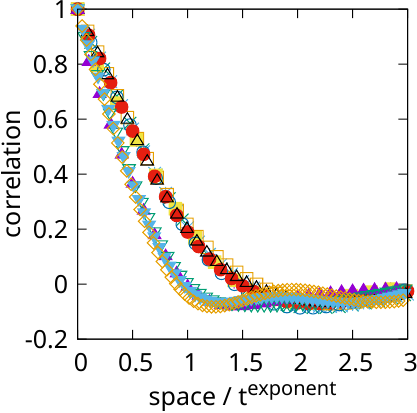}%
\hspace{0.01\columnwidth}%
\includegraphics[width=0.31\columnwidth,height=0.31\columnwidth]{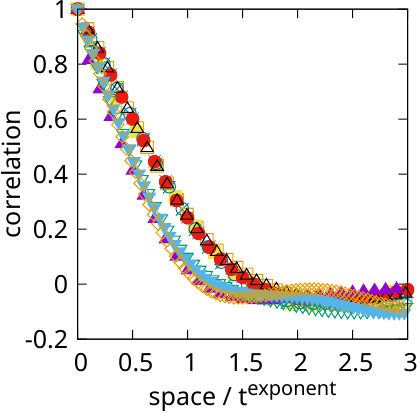}%
\hspace{0.01\columnwidth}%
\includegraphics[width=0.31\columnwidth,height=0.31\columnwidth]{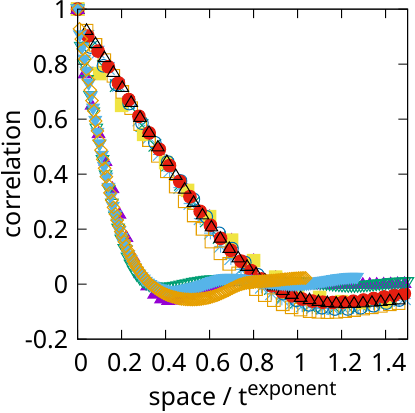}%
\\[0.05cm]
\caption{
Top row: characteristic domain size as a function of time
for initial passive fraction $c_0=0.8$. The three panels correspond,
from left to right, to evaporation probabilities
$\alpha=10^{-1}$, $10^{-2}$, and $10^{-3}$. In each panel,
data along the $x$, $y$, and $z$ directions are shown in
red, green, and blue, respectively, with $z$ the evaporation
direction. Solid lines indicate power-law references:
$0.12\,t^{1/4}$ and $0.8\,t^{1/3}$ (left),
$1.2\,t^{1/4}$ and $t^{1/3}$ (center),
$0.45\,t^{1/3}$ and $0.12\,t^{1/2}$ (right).
Bottom row: collapse of the two-point correlation functions
in the transverse and longitudinal directions for the same
parameters. For $\alpha=10^{-1}$ and $10^{-2}$ (left and center),
correlations are rescaled using exponents $1/4$ (transverse, 
upper set of curves)
and $1/3$ (longitudinal, lower set of curves), 
while for $\alpha=10^{-3}$ (right)
the rescaling uses $1/3$ and $1/2$, respectively.
Each collapse includes data at times
$t=10^3$, $5\times10^3$, $10^4$, and $1.5\times10^4$.
}
\label{fig:domain_scaling}
\end{figure}

\begin{figure}
\centering
\includegraphics[width=0.29\columnwidth,height=0.29\columnwidth]{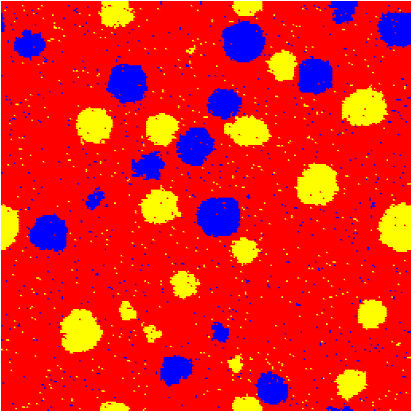}%
\hspace{0.03\columnwidth}%
\includegraphics[width=0.29\columnwidth,height=0.29\columnwidth]{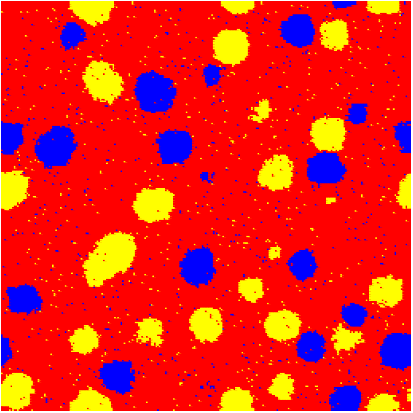}%
\hspace{0.03\columnwidth}%
\includegraphics[width=0.29\columnwidth,height=0.29\columnwidth]{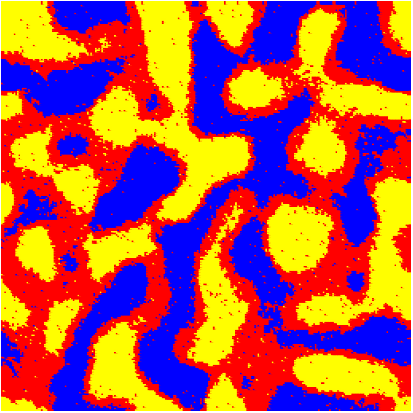}
\\[0.05cm]
\includegraphics[width=0.29\columnwidth,height=0.29\columnwidth]{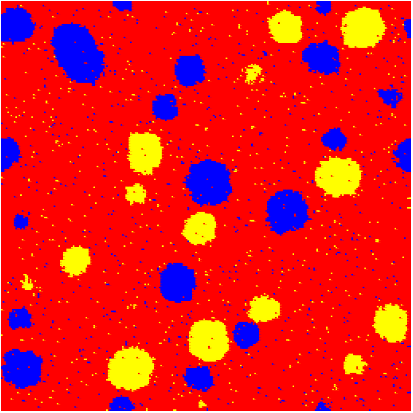}%
\hspace{0.03\columnwidth}%
\includegraphics[width=0.29\columnwidth,height=0.29\columnwidth]{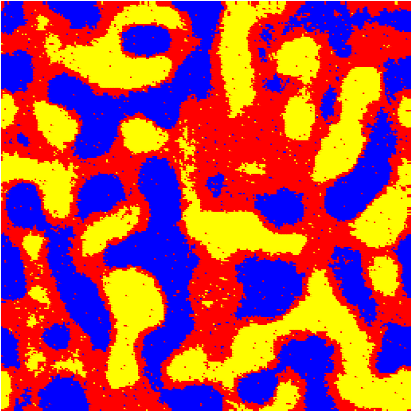}%
\hspace{0.03\columnwidth}%
\includegraphics[width=0.29\columnwidth,height=0.29\columnwidth]{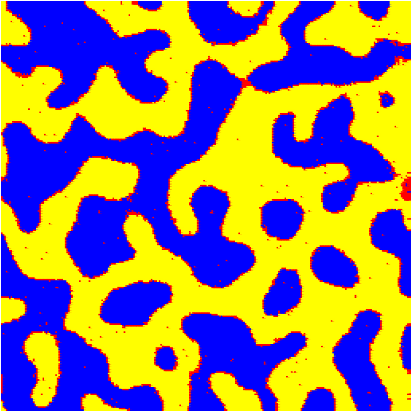}
\\[0.07cm]
\includegraphics[width=0.31\columnwidth,height=0.31\columnwidth]{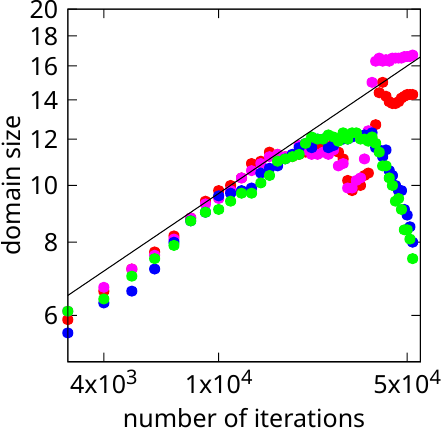}%
\hspace{0.01\columnwidth}%
\includegraphics[width=0.31\columnwidth,height=0.31\columnwidth]{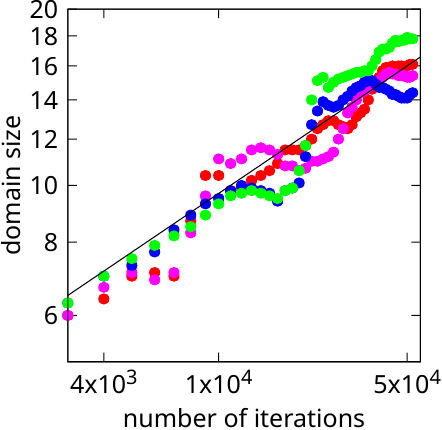}%
\hspace{0.01\columnwidth}%
\includegraphics[width=0.31\columnwidth,height=0.31\columnwidth]{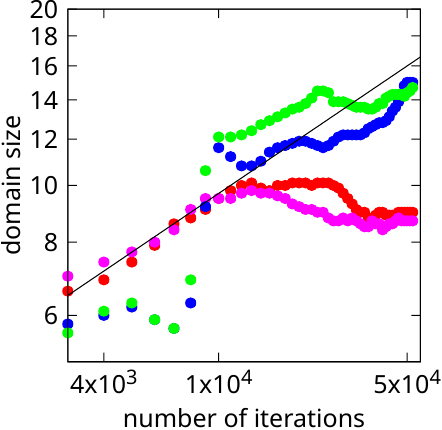}
\\[0.05cm]
\caption{Configuration at time 10000 on the planes 
$z=128$ (left), $z=192$ (center), and $z=224$ (right), 
$\alpha=10^{-3}$ (top row), and $\alpha=10^{-2}$ (center row). 
Bottom row:
characteristic domain size measured on the same plane as in 
the upper rows,
for 
$\alpha=10^{-2}$ in the $x$ direction (red), 
$\alpha=10^{-2}$ in the $y$ direction (magenta), 
$\alpha=10^{-3}$ in the $x$ direction (blue), 
$\alpha=10^{-3}$ in the $y$ direction (green).
Solid lines indicate the power-law reference
$0.45\,t^{1/3}$ in the three panels. 
}
\label{fig:sections}
\end{figure}

The quantitative analysis of domain growth, measured
through the two-point correlation function, is summarized
in Fig.~\ref{fig:domain_scaling}. The three panels correspond
to evaporation probabilities $\alpha=10^{-3}$, $10^{-2}$,
and $10^{-1}$, and report the characteristic domain size
along the $x$, $y$, and $z$ directions, with $z$ the
evaporation direction. For the slowest evaporation rate,
$\alpha=10^{-3}$, two clearly distinct regimes are observed:
the transverse directions follow the standard $t^{1/3}$
scaling of conserved dynamics, while the longitudinal
direction exhibits an extended $t^{1/2}$ growth regime.
For faster evaporation, $\alpha=10^{-2}$ and $10^{-1}$,
the anisotropy persists but the global transverse growth
displays an intermediate regime compatible with an
effective exponent close to $1/4$, before crossing over
toward diffusive-like behaviour, while the longitudinal
direction remains consistent with $t^{1/3}$.

The bottom row of Fig.~\ref{fig:domain_scaling} shows the
collapse of the two-point correlation functions for the same
parameters. Using the exponents extracted from the growth laws,
we observe an anisotropic collapse: correlations along
and perpendicular to the evaporation direction follow distinct
scaling, revealing that surface evaporation strongly breaks
the isotropy of spinodal decomposition.
For $\alpha=10^{-1}$ and $10^{-2}$, the transverse and
longitudinal correlations are consistently collapsed using
effective exponents $1/4$ and $1/3$, respectively,
whereas for $\alpha=10^{-3}$ the collapse
requires $1/3$ (transverse) and $1/2$ (longitudinal). 
This demonstrates that the global coarsening dynamics
remains anisotropic throughout the evolution.

Interestingly, when the analysis is restricted to a cubic region
far from the boundaries, the correlations follow the standard
Model~B isotropic scaling, indicating that the bulk preserves
the usual universal behavior. However, the anisotropy remains
the dominant feature when the full lattice is considered.

To investigate more closely how the boundary contribution combines
with the bulk dynamics, we performed a layer-wise analysis of the
lattice system. Figure~\ref{fig:sections} shows transverse sections at
$z=128$, $192$, and $224$, with the last plane being the closest to
the evaporating boundary. For $\alpha=10^{-3}$, the two innermost
sections are still rich in the passive component and consist mainly
of isolated, approximately circular domains of the two active
species. Closer to the boundary, at $z=224$, the active domains are
larger and start developing the interconnected morphology typical of
a nearly binary system, although a significant passive fraction is
still present. For $\alpha=10^{-2}$, the same morphological sequence
is observed at smaller distances from the center. In particular, the
section at $z=224$ already displays an almost binary morphology, with
the two active phases separated only by a thin layer of passive
sites.

The bottom row of Fig.~\ref{fig:sections} reports the characteristic
domain sizes measured along the $x$ and $y$ directions within each
section. The two transverse lengths remain comparable, as expected
from the symmetry parallel to the boundary. Despite the fluctuations
associated with measurements performed on a single $256\times256$
plane and for a single realization, the local growth is broadly
compatible with the Model~B law $t^{1/3}$. However, the crossover to
the late-time regime occurs at markedly different times in different
layers. At $z=128$ it occurs earlier for $\alpha=10^{-3}$, while at
$z=224$ it occurs earlier for $\alpha=10^{-2}$; at $z=192$, no clear
crossover is observed within the simulated time interval. Thus,
different transverse layers can simultaneously be at different
stages of the coarsening process.

\begin{figure}
\centering
\includegraphics[width=0.35\columnwidth,height=0.35\columnwidth]{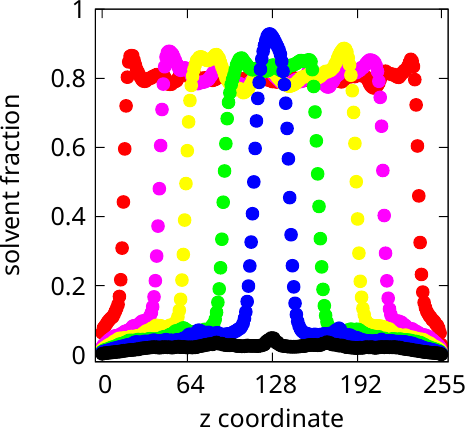}%
\hspace{0.1\columnwidth}%
\includegraphics[width=0.35\columnwidth,height=0.35\columnwidth]{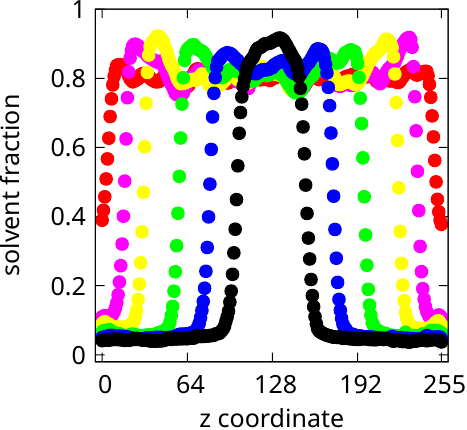}%
\caption{Fraction of passive conponent as function of the $z$ coordinate 
for  
$\alpha=10^{-2}$ (left) and $\alpha=10^{-3}$ (right) after 
$10^3$ (red), 
$5\times10^3$ (magenta), 
$10^4$ (yellow), 
$2\times10^4$ (green), 
$3\times10^4$ (blue), 
$4\times10^4$ (black)
iterations.
}
\label{fig:passive_density}
\end{figure}

This interpretation is supported by the passive-component profiles
shown in Fig.~\ref{fig:passive_density}. The passive fraction is
largest around the central plane and remains approximately uniform
over an inner region, before rapidly decreasing toward the
evaporating boundaries. Equivalently, the total concentration of the
two active species, given by one minus the passive fraction, increases
sharply toward the boundaries. As evaporation proceeds, the central
plateau becomes progressively narrower, and this reduction is faster
for $\alpha=10^{-2}$ than for $\alpha=10^{-3}$. The boundary-induced
composition gradient therefore propagates through the system and
causes different layers to undergo the transition from a
passive-rich morphology to an almost binary one at different times.

The global correlation functions combine contributions from all
these dynamically different layers. In particular, for sufficiently
fast evaporation, the progressive crossover of successive sections
can produce an effective transverse growth exponent smaller than the
local Model~B value, providing a possible explanation for the
intermediate exponent observed for $\alpha=10^{-2}$ and
$\alpha=10^{-1}$ in Fig.~\ref{fig:domain_scaling}. The layer-wise
analysis therefore indicates that this effective behavior does not
necessarily reflect a distinct local coarsening law, but can arise
from the spatially inhomogeneous propagation of the
evaporation-induced composition profile.

\begin{figure}
\centering
\includegraphics[width=0.31\columnwidth,height=0.31\columnwidth]{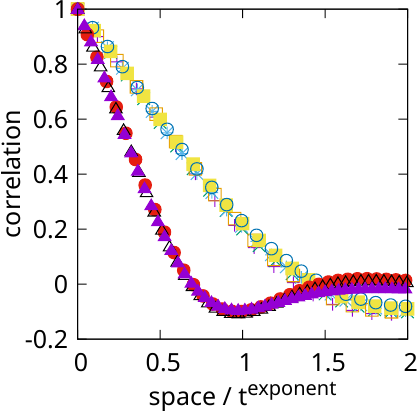}%
\hspace{0.01\columnwidth}%
\includegraphics[width=0.31\columnwidth,height=0.31\columnwidth]{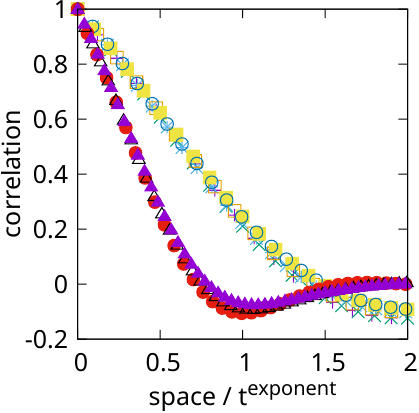}%
\hspace{0.01\columnwidth}%
\includegraphics[width=0.31\columnwidth,height=0.31\columnwidth]{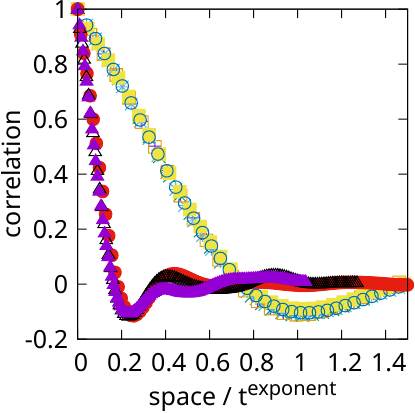}%
\\[0.05cm]
\caption{Collapsed two-point correlation functions for $c_0 = 0.4$ in the
lattice model. From left to right: evaporation probabilities $\alpha$ and 
time exponent as in
the $c_0 = 0.8$ figure. Four times are
used for the collapse in each panel: $t=1000, 5000, 10000, 15000$ iterations.
}
\label{fig:c04_collapse}
\end{figure}

These results show that boundary conditions alone can
control coarsening laws without altering bulk universality.
To test the robustness of the anisotropic coarsening, we consider
a lower initial passive fraction, $c_0 = 0.4$. Figure~\ref{fig:c04_collapse}
shows the collapsed two-point correlation functions along all
three spatial directions for the same evaporation probabilities as
for $c_0 = 0.8$. In all cases, the longitudinal and transverse
correlations collapse onto direction-dependent master curves,
demonstrating that the anisotropic scaling scenario is fully
preserved even at lower passive component content.
The extracted exponents are consistent with those
observed for $c_0 = 0.8$, suggesting that the anisotropic
growth laws are robust with respect to variations in the
initial passive fraction. Similar results for $L=128$
further indicate that the observed anisotropy is not due
to finite-size effects, while bulk regions away from the
boundaries continue to display standard isotropic
scaling.

\begin{figure}[t]
\centering
\includegraphics[width=0.31\columnwidth,height=0.31\columnwidth]{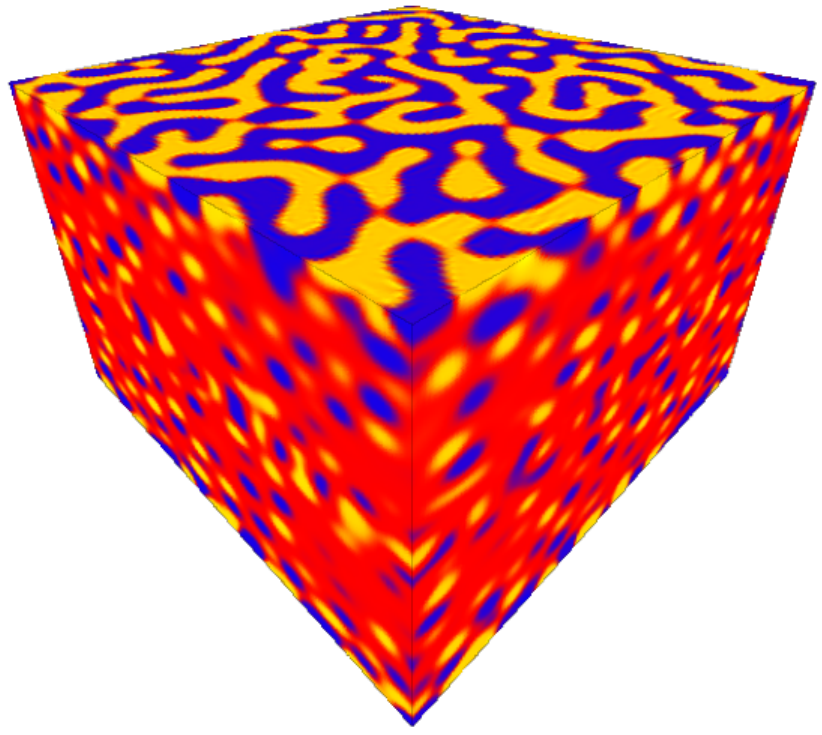}%
\hspace{0.01\columnwidth}%
\includegraphics[width=0.31\columnwidth,height=0.31\columnwidth]{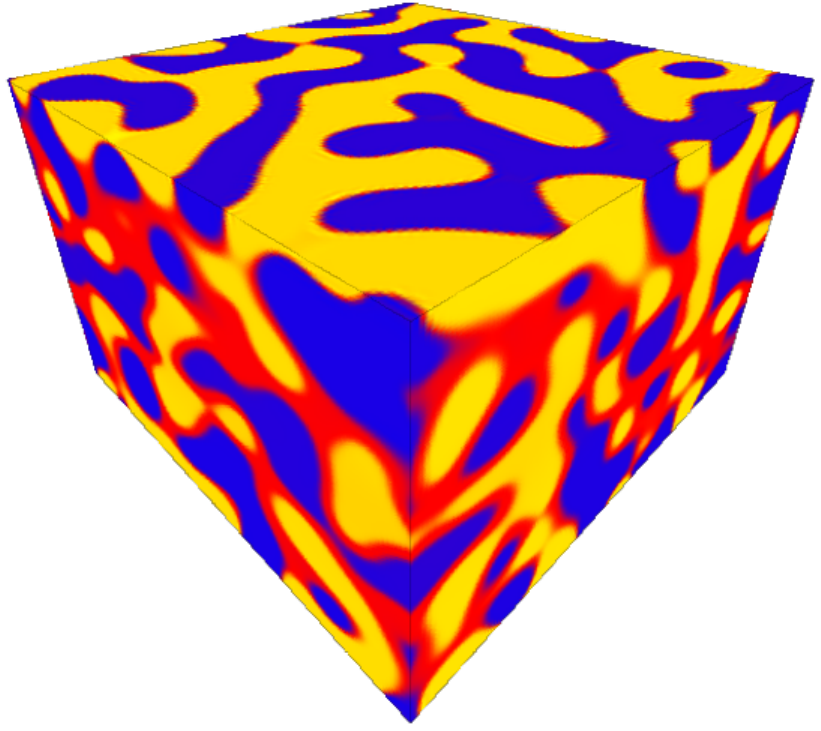}%
\hspace{0.01\columnwidth}%
\includegraphics[width=0.31\columnwidth,height=0.31\columnwidth]{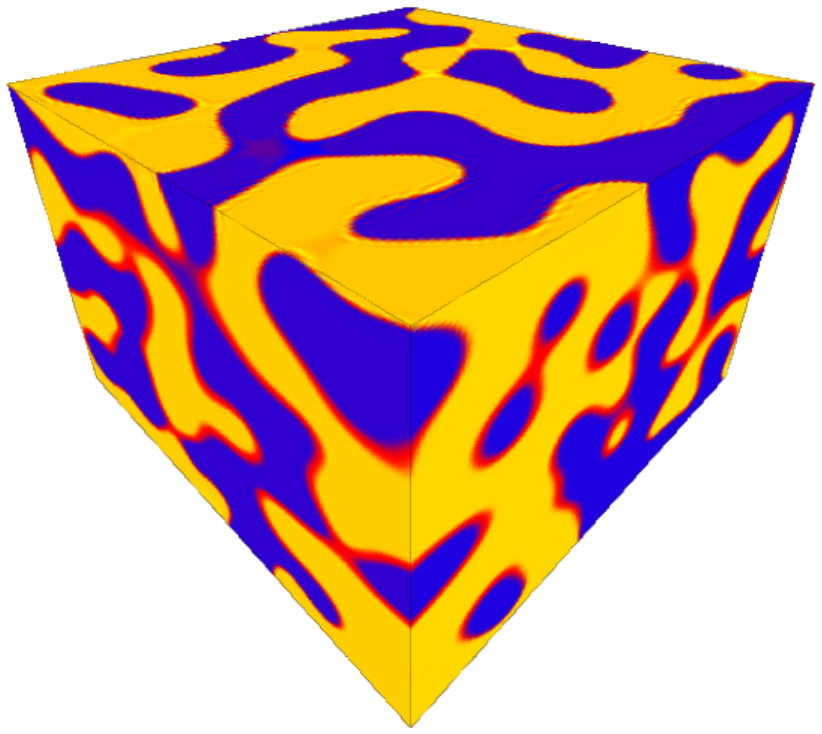}
\\[0.05cm]
\caption{
Typical configurations of the continuum model 
in presence of evaporation with 
$\beta=1$,
$1-\phi_0=0.8$,
$\alpha'=100$
at time $0.1$, $1.7$, $3.5$
(left to right). 
}
\label{fig:continuous_model_conf}
\end{figure}

\begin{figure}[t]
\centering
\includegraphics[width=0.31\columnwidth,height=0.31\columnwidth]{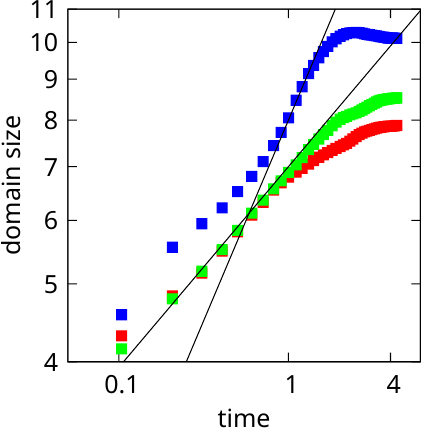}%
\hspace{0.01\columnwidth}%
\includegraphics[width=0.31\columnwidth,height=0.31\columnwidth]{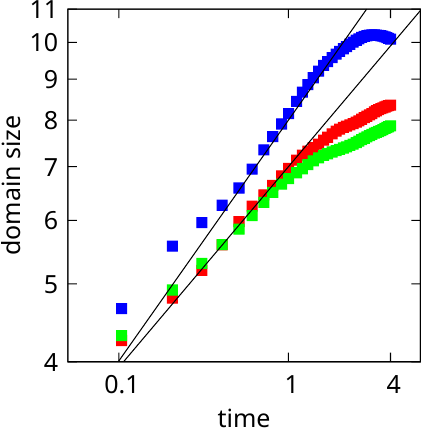}%
\hspace{0.01\columnwidth}%
\includegraphics[width=0.31\columnwidth,height=0.31\columnwidth]{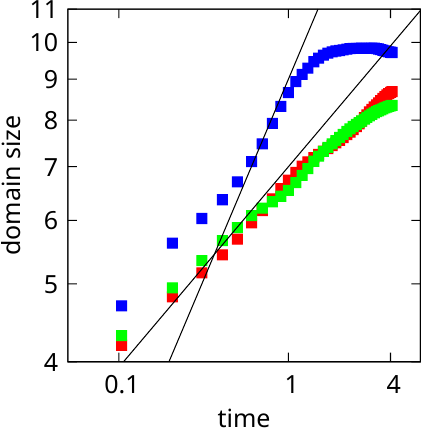}%
\\[0.05cm]
\caption{
Characteristic domain size as a function of time
for initial passive fraction $1-\phi_0=0.8$,
$\beta=1$,
and
$\alpha'=1,10,100$
(left to right). Data along
the $x$, $y$, and $z$ directions are shown in
red, green, and blue, respectively, with $z$ the evaporation
direction. Solid lines indicate power-law references:
$7\,t^{1/4}$ and $8\,t^{1/2}$ (left),
$7\,t^{1/4}$ and $8\,t^{1/3}$ (center),
$7\,t^{1/3}$ and $9\,t^{1/2}$ (right).
}
\label{fig:continuous_model}
\end{figure}

We observe qualitatively similar morphologies and
quantitatively consistent anisotropic growth in the
continuum model, suggesting that the surface-driven
anisotropy is not a lattice artifact.
The morphology evolution is illustrated in
Fig.~\ref{fig:continuous_model_conf}, where three cubic
snapshots show the elongation of domains along the
evaporation direction. The transverse and longitudinal
domain sizes are reported in Fig.~\ref{fig:continuous_model}.
The data indicate that the transverse growth in the $x$
and $y$ directions remains comparable, while the
longitudinal growth along $z$ is systematically faster.
This behavior is consistent with the surface-driven
anisotropic coarsening observed in the lattice model,
supporting the robustness of the effect beyond the
lattice description.

A simple scaling argument may explain the 
anisotropic coarsening. While in directions parallel to the
boundary, the system 
evolves via Model B dynamics,
along the evaporation direction the boundary acts directly
on the passive component breaking
global mass conservation and generating a macroscopic
gradient in the passive concentration, which is expected to
drive a net transport across the system.
The resulting transport affects the dynamics of the active
components through their coupling to the passive one, so 
active species follow the large-scale transport
induced by the boundary condition.
In the absence of advective mechanisms, this transport is
expected to be governed by diffusion, which controls the
relaxation of the boundary-induced gradient. Over a length
scale $\ell(t)$, this leads to the characteristic scaling
$\ell(t)\sim (Dt)^{1/2}$.
Thus, 
the observed anisotropy thus appears to result from the
interplay between curvature-driven coarsening in the bulk
and a boundary-induced transport mechanism.

This simple argument is expected to apply most directly in the
weak-evaporation regime. For larger evaporation rates, the
layer-wise analysis presented above indicates that different
regions of the sample undergo the morphological transition at
different times, so that the global dynamics cannot be
described by a single coarsening length. In this case, the
observed intermediate exponent should be regarded as an
effective global behavior arising from the superposition of
different local coarsening stages.

In conclusion, we have shown that surface evaporation can
induce anisotropic coarsening in ternary mixtures of two
active and one passive components, leading to a departure
from isotropic coarsening at the system level. 
Domain growth becomes direction-dependent, with different
effective scaling behavior parallel and perpendicular to
the mass-loss direction.
Interestingly, bulk regions away from the boundaries
appear to retain standard Model B scaling, suggesting
that the observed anisotropy originates from
boundary-driven concentration gradients rather than from
modifications of the intrinsic bulk dynamics.
Evaporation thus provides a minimal mechanism for
symmetry breaking in conserved phase separation,
enabling direction-dependent scaling without altering
bulk universality.
This identifies boundary forcing as a generic route to
anisotropic scaling in conserved systems.

\acknowledgments
NJ and AM are involved in Swedish Energy Agency's project Solar 
Electricity Research Centre (SOLVE) with grant number 52693-1. 
The computations were enabled by resources provided by the National 
Academic Infrastructure for Supercomputing in Sweden (NAISS), partially 
funded by the Swedish Research Council through grant agreement no. 2022-06725,
as well as by the SBAI Department (Sapienza).
ENMC thanks the PRIN 2022 project
``Mathematical Modelling of Heterogeneous Systems (MMHS)",
financed by the European Union - Next Generation EU,
CUP B53D23009360006, Project Code 2022MKB7MM, PNRR M4.C2.1.1.
ENMC thanks, also, the Mathematics Department of the Karlstad University 
for warm hospitality and GNFM--INDAM.



\bibliographystyle{unsrt}
\bibliography{morpho}

\end{document}